\documentclass[%
 reprint,
 amsmath,amssymb,
 aps,
 nourl,
]{revtex4-2}

\usepackage{mathrsfs}
\usepackage{xcolor}
\usepackage{graphicx}
\usepackage{dcolumn}
\usepackage{bm}

\usepackage[hidelinks]{hyperref}

\begin{document}

\preprint{APS/123-QED}

\title{From Quantum Relative Entropy to the Semiclassical Einstein Equations}

\author{Philipp Dorau}
\email{philipp.dorau@uni-leipzig.de}
\author{Albert Much}%
\email{much@itp.uni-leipzig.de}
\affiliation{Institut f\"ur Theoretische Physik, Universit\"at Leipzig, \\
Br\"uderstraße 16, 04103 Leipzig, Germany}

\date{March 2, 2026}

\begin{abstract}
We provide arguments indicating that the semiclassical Einstein equations follow from quantum relative entropy and its proportionality to an area variation. Using modular theory, we establish that the relative entropy between the vacuum state and coherent excitations of a scalar quantum field on a bifurcate Killing horizon is given by the energy flux across the horizon. Under the assumption of the Bekenstein-Hawking entropy-area formula, this energy flux is proportional to a variation in the surface area of the horizon cross section. The semiclassical Einstein equations follow automatically from this identification. Our approach provides a quantum field theoretic generalization of Jacobson's thermodynamic derivation of the Einstein equations, replacing classical thermodynamic entropy with the well-defined quantum relative (Araki-Uhlmann) entropy. This suggests that quantum information plays a central role in what is often seen as a zeroth order approximation of a theory of quantum gravity, namely quantum field theory in curved spacetimes.
\end{abstract}

\maketitle


\paragraph*{Introduction.}
Jacobson illustrates in Ref. \cite{Jacobson:1995eos} that the Einstein equations can be derived from the thermodynamic relation $\delta Q = T \delta S$ applied to local Rindler horizons, where the heat flux through the horizon is connected to the entropy change, itself proportional via the Bekenstein-Hawking formula \cite{Bekenstein:1973bhe,Hawking:1975rad,Hawking:1976bhe} to the variation of the horizon area. In his argument, he invokes the Hawking effect \cite{Hawking:1975rad} and, most notably, the Unruh effect \cite{Unruh:1976effect}, both of which fall within the domain of quantum field theory (QFT). This naturally raises the question: Can an analogous derivation be formulated entirely within QFT?

Various approaches have already linked semiclassical gravity to entropic and thermodynamic properties of quantum fields. Most notably, Jacobson proposed a route employing the entanglement entropy of causal diamonds \cite{Jacobson_2016}, related to previous results in \cite{Faulkner_2014}, see also \cite{Ent1, Ent2, Ent3} for related monographs. More recently, connections between the thermodynamics of stretched light cones and gravitational dynamics have been developed \cite{ALP:2025scstt, AGLL:2025gfet}, and Jacobson’s original work has been extended within the framework of causal fermion systems, see \cite{CFI:2020jcfs}.

In this work, we formulate a quantum field theoretic extension of Jacobson’s argument based on the relative entropy, which quantifies the distinguishability of quantum states and plays a central role in quantum information theory, see \cite{Longo:2019relent,Hollands:2020relent,HI:2019new,Casini:2019qst}. We thereby follow Casini \cite{Casini:2008rebb}, who identified the relative entropy as the natural entropic quantity for a quantum field theoretic derivation of the Bekenstein bound, which is classically derived from the black hole area law \cite{Casini:2008rebb}. Furthermore, the commonly employed von Neumann entropy is typically ill defined in QFT due to ultraviolet divergences arising from the type $\mathrm{III}$ structure of local von Neumann algebras, which is a manifestation of vacuum fluctuations \footnote{More precisely, the von Neumann entropy is ill defined for local QFT algebras because sharp localization yields type III von Neumann algebras with infinitely many UV‑entangled modes across any boundary, so that neither a trace nor a density matrix exists to make the entropy finite.}. In contrast, the relative entropy \footnote{For example, in the simpler case of quantum mechanics (type $\mathrm{I}$ algebras), the relative entropy is defined in terms of density operators $\rho,\varrho$ as $S^\mathrm{rel}(\rho\Vert\varrho) = \mathrm{Tr} (\rho\log\rho)-\mathrm{Tr} (\rho\log\varrho )$.} remains well defined and, in our present setting, finite. We also refer to \cite{Bianconi:2025gfe, Bianconi_2025} for a notable approach relating gravity to quantum relative entropy.

Our strategy is as follows. Invoking the equivalence principle, we approximate any sufficiently small spacetime region by Minkowski space and consider a uniformly accelerated observer associated with a local Rindler horizon. This setting provides a local approximation of a bifurcate Killing horizon, on which we compute the relative (Araki-Uhlmann) entropy, see \cite{Araki:1976relent,Uhlmann:1977relent}, between the vacuum and a coherent excitation of a Klein-Gordon field by using modular theory, see \cite{Takesaki:1970tomita}. The resulting expression is given in terms of the expectation value of the field’s energy momentum tensor (cf. \cite{Casini:2019qst,KPV:2021ea}), which is, in turn, directly related to the energy flux across the horizon \cite{KW:1991hadamard}. Following the reasoning of Ref. \cite{Jacobson:1995eos}, the proportionality between the relative entropy and the area variation of the horizon cross section then recovers the semiclassical Einstein equations.

\paragraph*{Local Spacetime Geometry.}
Consider a spacetime manifold $\mathcal{M}$, together with a sufficiently small neighborhood $\mathcal{U}\subset\mathcal{M}$ around some point $p\in\mathcal{M}$ such that its causal completion $\overline{\mathcal{U}}^\diamond$ is globally hyperbolic, and, following the equivalence principle \cite{Wald:1984gr}, the spacetime metric $g_{ab}$ restricted to $\mathcal{U}$ is well approximated by that of a local inertial frame, i.e.,
\begin{equation}\label{LocalApproximation}
  g_{ab}\big\vert_{\mathcal{U}} \approx \eta_{ab}
\end{equation}

\noindent where $\eta_{ab}$ denotes the flat Minkowski metric. 

\begin{figure}[h!]
\centering
\includegraphics[width=0.41\textwidth]{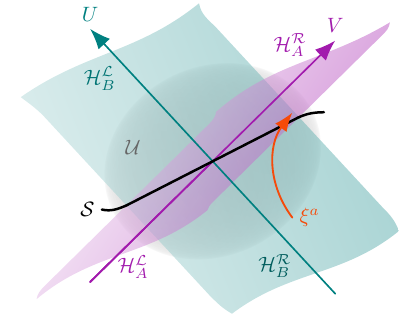}
\caption{Sketch of a local region $\mathcal{U}\subset\mathcal{M}$, faintly shaded in gray, endowed with a local Rindler horizon, consisting of the null hypersurfaces $\mathcal{H}_A$ and $\mathcal{H}_B$ intersecting at the horizon cross section $\mathcal{S}$. Within $\mathcal{U}$, $\mathcal{H}_A$ is affinely parametrized by $V$, and likewise $\mathcal{H}_B$ by $U$, yielding a local double null coordinate system. Consequently, the region $\mathcal{U}$ is separated into two wedge-shaped regions, correspondingly decomposing $\mathcal{H}_A,\mathcal{H}_B$ into $\mathcal{H}_A^{\mathcal{R}},\mathcal{H}_A^{\mathcal{L}}$ and $\mathcal{H}_B^{\mathcal{L}},\mathcal{H}_B^{\mathcal{R}}$, respectively. The orange curve indicates the flow generated by the local approximate boost Killing vector field $\xi^a$, which becomes null on $\mathcal{H}_A$ and $\mathcal{H}_B$.}
\label{FigureBifurcateKillingHorizon}
\end{figure}

In this locally flat region, the generating vector field $\xi^a$ of Lorentz boosts constitutes an approximate Killing vector field on $\mathcal{U}$. Moreover, $\xi^a$ becomes null on two null hypersurfaces $\mathcal{H}_A$ and $\mathcal{H}_B$, to which it is both tangent and normal, and which intersect on a spacelike $2$–surface $\mathcal{S}$, as illustrated in Fig. \ref{FigureBifurcateKillingHorizon}. Hence, the region $\mathcal{U}$ possesses an approximate bifurcate Killing horizon, see \cite{KW:1991hadamard,SV:1996tomitaki}, namely the local Rindler horizon employed in \cite{Jacobson:1995eos}.

\paragraph*{Algebraic QFT and Relative Entropy.} 
Next, we  turn to the corresponding algebraic formulation of a local QFT. Consider a real, minimally coupled scalar field $\Phi$ of mass $m$ on a globally hyperbolic spacetime region $(\overline{\mathcal{U}}^\diamond,g)$ satisfying the Klein-Gordon equation
\begin{equation}\label{KleinGordonEquation}
\left( \square_{g} + m^2 \right) \Phi = 0.
\end{equation}
\noindent Algebraic quantization of $\Phi$ then gives rise to the canonical commutation relations algebra $\mathscr{C}$ of scalar field operators, see, e.g., \cite{FR:2019aqft}.

Let $\tau^t$ denote the flow generated by the (approximate) Killing vector field $\xi^a$, and let $\omega_0$ be a quasifree $\tau^t$-invariant Hadamard state on $\mathscr{C}$. This class of states forms the natural curved spacetime generalization of the Minkowski vacuum \cite{KW:1991hadamard}. In the specific case of a Rindler horizon in flat spacetime, the relevant $\tau^t$-invariant state is indeed the Minkowski vacuum itself, i.e., the unique Poincaré-invariant Hadamard state, see \cite{Haag:1996lqp}.

Kay and Wald \cite{KW:1991hadamard} derived that the restriction of any such $\omega_0$ to the Killing horizon $\mathcal{H}_B$ associated to $\xi^a$ possesses the universal scaling limit two-point function
\begin{equation}\label{ScalingLimitTwoPointFunction}
\Lambda(\phi,\psi) = -\frac{1}{\pi}\int \frac{\phi(U,s)\,\psi(U',s)}{(U-U'-i0^+)^2}\, dU\, dU'\, d\mathrm{vol}_\mathcal{S},
\end{equation}

\noindent see also \cite{HNS:1984scalim}, where $\phi,\psi \in C_c^\infty(\mathcal{H}_B)$ are compactly supported solutions of \eqref{KleinGordonEquation} restricted to the horizon $\mathcal{H}_B$, $U$ denotes a null coordinate affinely parametrizing $\mathcal{H}_B$, as before, and $s$ denotes coordinates on the bifurcation surface $\mathcal{S}$. In particular, this two-point function satisfies the Kubo-Martin-Schwinger (KMS) condition \footnote{States satisfying the KMS condition generalize the notion of Gibbs ensembles beyond finite systems, see \cite{KMS1, KMS2, KMS3}.} at inverse temperature $\beta = \tfrac{2\pi}{\kappa}$ with respect to the projected Killing flow $\left.\tau^t\right\vert_{\mathcal{H}}$ \cite{KW:1991hadamard}, thereby recovering the Hawking and, in the Rindler case, the Unruh temperature. Note that, in contrast to stationary black holes, the boost Killing field generating a Rindler horizon has no unique normalization, since any rescaling changes the proper acceleration $a$ of the corresponding uniformly accelerated observers and thereby the surface gravity $\kappa$ of the associated horizon. 

Building on these results, Summers and Verch \cite{SV:1996tomitaki} reformulated the thermal properties of QFT on bifurcate Killing horizons in a purely operator algebraic language, using Tomita-Takesaki modular theory, see \cite{Takesaki:1970tomita,Borchers:2000modth}. Let $\mathscr{A}_\mathcal{R}$ denote the von Neumann algebra of observables localized in the right wedge of a bifurcate Killing horizon, see Fig. \ref{FigureBifurcateKillingHorizon}, and let $\omega_0$ be a KMS state on $\mathscr{A}_\mathcal{R}$ with respect to the Killing flow $\tau^t$. Then, there exists a subalgebra $\mathscr{N}_\mathcal{R} \subset \mathscr{A}_\mathcal{R}$, localized on the horizon portion $\mathcal{H}_B^{\mathcal{R}}$, such that the restricted state $\left.\omega_0\right\vert_{\mathscr{N}_\mathcal{R}}$ is a KMS state for the projected Killing flow $\left.\tau^t\right\vert_{\mathcal{H}}$ at inverse temperature $\beta = \frac{2\pi}{\kappa}$ \cite{SV:1996tomitaki}. In the corresponding Gelfand-Naimark-Segal (GNS) representation $(\mathscr{H}_0, \pi_0, \Omega_0)$ consisting of a representation $\pi_0$ of $\mathscr{N}_\mathcal{R}$ on the Hilbert space $\mathscr{H}_0$, and a cyclic vector $\Omega_0\in\mathscr{H}_0$, see \cite{GN:1943gns,Segal:1947gns}, $\Omega_0$ is separating for $\mathscr{N}_\mathcal{R}$, and the associated modular group $(\Delta_\mathcal{R}^{it})_{t \in \mathbb{R}}$ acts geometrically as affine dilations $\mathfrak{D}_{2\pi t}$ along $\mathcal{H}_B^\mathcal{R}$ \cite{SV:1996tomitaki}, i.e.,
\begin{equation}\label{ModularOperatoR}
\Delta_\mathcal{R}^{it} = \mathfrak{D}_{2\pi t}.
\end{equation}

\noindent This result generalizes the Bisognano-Wichmann property of the Minkowski vacuum, see \cite{BW:1975thm}, to any bifurcate Killing horizon and provides a quantum information theoretic derivation of the Unruh and Hawking temperatures of bifurcate Killing horizons.

Following Ref. \cite{KPV:2021ea}, the horizon algebra $\mathscr{N}_\mathcal{R}$ is constructed explicitly in terms of the Weyl algebra $\mathscr{W}_\mathcal{R}$ associated to the symplectic space $\left(C_c^\infty(\mathcal{H}_B^\mathcal{R}), \sigma \right)$ of test functions supported on the right horizon portion $\mathcal{H}_B^\mathcal{R}$ with symplectic form
\begin{equation}\label{SymplecticForm}
\sigma(\phi,\psi) = 2\mathrm{Im}\Lambda(\phi,\psi) = \int \big(\phi (\partial_U\psi) - \psi (\partial_U\phi)\big) \,dU d\mathrm{vol}_\mathcal{S},
\end{equation}
as follows. Recall that $\mathscr{W}_\mathcal{R}$ is generated by unitaries $W(\phi)\in\mathscr{W}_\mathcal{R}$ ($\phi\in C_c^\infty(\mathcal{H}_B^\mathcal{R})$) that fulfill the Weyl relations
\begin{align}
W(\phi)^\dagger &= W(-\phi),\\
W(\phi)W(\psi)&= e^{-\tfrac{i}{2}\sigma(\phi,\psi)} W(\phi + \psi).
\end{align}

\noindent The von Neumann algebra $\mathscr{N}_\mathcal{R}$ is then obtained by taking the double commutant of the vacuum representation $\pi_0$ of $\mathscr{W}_\mathcal{R}$, i.e., $\mathscr{N}_\mathcal{R} = \pi_0 \left( \mathscr{W}_\mathcal{R} \right)''$ \cite{KPV:2021ea}. Furthermore, in this representation \footnote{For improved readability, we adopt the mild abuse of notation that $W$ and $\Phi$ are implicitly understood in their appropriate representation, see \cite{FR:2019aqft} for details.}, the Weyl operators are identified with exponentials of the quantized field, i.e., $W(\phi)\cong e^{i\Phi(\phi)}$.

On $\mathscr{W}_\mathcal{R}$, the quasifree state $\omega_0$ is induced by the scaling limit two-point function $\Lambda$ via
\begin{equation}
    \omega_0\big(W(\psi)\big) = e^{-\tfrac{1}{2}\Lambda(\psi,\psi)},
\end{equation} 

\noindent and its coherent excitations on $\mathscr{W}_\mathcal{R}$ are defined by
\begin{equation}
    \omega_\phi\big(W(\psi)\big) = \omega_0 \big( W(\phi)^\dagger W(\psi) W(\phi)\big),
\end{equation}

\noindent for some $\phi\in C_c^\infty(\mathcal{H}_B^\mathcal{R})$, so that the corresponding GNS vectors $\Omega_0,\Omega_\phi$ are related via $\Omega_\phi = e^{i\Phi(\phi)}\Omega_0$ \cite{KPV:2021ea}.

Hence, we have gathered all necessary ingredients to explicitly compute the relative entropy between the state $\omega_0$ and its coherent excitation $\omega_\phi$ by using the Araki-Uhlmann formula for coherent states \footnote{Note that for noncoherent states, Formula \eqref{ArakiUhlmannFormula} does \emph{not} hold, so that the expression for the relative entropy generally becomes significantly more involved.} \cite{Araki:1976relent,Uhlmann:1977relent,HI:2019new,KPV:2021ea}
\begin{equation}\label{ArakiUhlmannFormula}
    S^{\mathrm{rel}}(\omega_0\Vert\omega_\phi)= i\left.\frac{d}{dt}\right\vert_{t=0}\langle\Omega_\phi \vert \Delta^{it}_\mathcal{R}\Omega_\phi\rangle.
\end{equation}

\noindent Using the geometric action \eqref{ModularOperatoR} of the modular operator, and repeating the calculations in Ref. \cite{KPV:2021ea}, we hence obtain that the relative entropy only depends on the symplectic form \eqref{SymplecticForm} via
\begin{equation}
S^\mathrm{rel}\left(\omega_0 \Vert \omega_\phi \right) = \frac{1}{2} \left.\frac{d}{dt}\right\vert_{t=0}\sigma(\phi^t,\phi),
\end{equation}

\noindent where $\phi^t (U,s):= (\mathfrak{D}_{2\pi t} \phi)(U,s) = \phi(e^{2\pi t} U,s)$. Ultimately, a direct computation yields that the relative entropy takes the form \cite{KPV:2021ea}
\begin{equation}\label{RelativeEntropy}
S^\mathrm{rel}\left(\omega_0 \Vert \omega_\phi \right) = -2\pi \int_{\mathcal{H}_B^\mathcal{R}} U (\partial_U \phi)^2 dU d\mathrm{vol}_\mathcal{S}.
\end{equation}

In particular, the relative entropy admits the reformulation \footnote{In fact, the relative entropy can be interpreted as a Noether charge. This was explicitly proven for de Sitter spacetime in \cite{FMP:2023ren} and the argument extends to the case at hand. We also refer to \cite{IW:1994ent} for a broader discussion.}
\begin{align}\label{RelativeEntropyEnergyRelation}
    S^\mathrm{rel}\left(\omega_0 \Vert \omega_\phi \right) = -{2\pi}\int_{\mathcal{H}_B^\mathcal{R}} U \langle :T_{ab}:\rangle_{\omega_\phi}\xi^a\xi^bdU d\mathrm{vol}_\mathcal{S}.
\end{align}

\noindent To see this, let $:T_{ab}:$ denote the normal ordered \footnote{Here, normal ordering is understood in the sense of Hadamard regularization.} energy momentum tensor for the quantized field $\Phi$, given by
\begin{align}
    :T_{ab}:\;=\;:\nabla_{a}\Phi\nabla_{b}\Phi:-\frac{1}{2}g_{ab} (m^2:\Phi^2:+:\nabla_{c} \Phi \nabla^{c} \Phi:).
\end{align}

\noindent Using the metric's double null structure
\begin{equation}
ds^2 = -2A(U,V) \,dUdV + h_{ij} \, dx^i dx^j,
\end{equation}

\noindent the expectation value of the energy density $:T_{UU}:$ in the coherent state $\omega_{\phi}$ then reads \footnote{We use the fact that since $W$ implements  linear canonical  transformations on the field,  one thus has $W :O: W^{\dagger} = \; : W O W^{\dagger}:$. }  
\begin{align}
        \langle :T_{UU}:\rangle_{\omega_\phi}&=      \langle\Omega_0\vert e^{-i\Phi(\phi)} :\partial_{U}\Phi(U,s)^2:e^{i\Phi(\phi)}\Omega_0\rangle \\
        &=    \langle\Omega_0\vert :(e^{-i\Phi(\phi)}\partial_{U}\Phi(U,s)e^{i\Phi(\phi)})^2:\Omega_0\rangle\nonumber .
\end{align}

\noindent By taking into account the unitary transformation of a field with respect to Weyl operators $W(\phi)\Phi(x)W(-\phi)=\Phi(x)+\phi(x)$, see \cite[Eq. (3.3)]{HI:2019new}, \cite[Prop. 140]{combescure2012coherent}, or \cite[Eq. (5.18)]{Degner2013}, the former expression further reduces to
\begin{align}
\langle\Omega_0\vert :(\partial_{U}\Phi(U,s)-\partial_{U}\phi)^2:\Omega_0\rangle =(\partial_{U}\phi)^2.
\end{align}

\noindent Hence, the expectation value of the energy density $:T_{UU}:$ in the coherent state $\omega_\phi$ is
\begin{align}
    \langle :T_{ab}:\rangle_{\omega_\phi}\xi^a\xi^b=  (\partial_{U} \phi)^2.
\end{align} 

\noindent This expression is equal to the energy density $T^\mathrm{cl}_{UU}$ of a classical solution $\phi$, see \cite[Eqs. (6.7)  and (6.37)]{KW:1991hadamard}. We also refer to \cite[Eq. (1.18)]{Casini:2019qst} and \cite[Eq. (74)]{KPV:2021ea} for related arguments. Most importantly, this correspondence establishes an identification between the relative entropy and the energy flux along the Killing flow through $\mathcal{H}_B^\mathcal{R}$ as discussed in \cite[Sec. 6.4]{KW:1991hadamard}, see also \cite{HI:2019new,KPV:2021ea,DAngelo:2021bhre}. This connection is particularly significant because it provides a quantum field theoretic formulation of the energy flux $\delta Q$ that underlies Jacobson's thermodynamic derivation of the Einstein equations \cite{Jacobson:1995eos}.

\paragraph*{The Semiclassical Einstein Equations.}
Having established the equality between the relative entropy and the energy flux across a local Rindler horizon, Jacobson’s argument \cite{Jacobson:1995eos} applies: The flux $\delta Q$ is proportional to the horizon entropy variation, which is, in turn, proportional to the variation $\delta A$ of the surface area of the horizon cross section $\mathcal{S}$. This is indeed consistent with Ref. \cite{KPV:2021ea}, where the relative entropy between coherent excitations is proportional to the surface area $A(\mathcal{O}) \leq A(\mathcal{S})$ of a local patch $\mathcal{O} := \mathrm{supp}(f)\cap\mathcal{S} \subset \mathcal{S}$ of the cross section of any spherically symmetric future outer trapping horizon, and with Ref. \cite{DFGMMP:2024dS}, where the relative entropy between coherent states on de Sitter horizons is directly proportional to the average variation of the respective horizon cross section area. In light of this, we formulate the relation between the information theoretic energy flux and the geometric area variation more precisely as follows.

On a local Rindler horizon, we begin with the proportionality between the relative entropy \eqref{RelativeEntropy} for coherent excitations and an area variation $\delta A$ of the horizon cross section \footnote{Although the Rindler horizon cross section has infinite surface area, one can still meaningfully define its variation $\delta A$, as in Jacobson’s original argument \cite{Jacobson:1995eos}. We also refer to \cite{BS:2010ar} for a related discussion yielding an area variation closely aligned with Eq. \eqref{AreaVariationEntropicEnergy}.}. Furthermore, $\delta A$ can be modeled by a linear perturbation $\tilde{h}_{ij}$ of the induced metric $h_{ij}$ on $\mathcal{S}$, sourced by the energy content of the coherent excitation as quantified by the relative entropy \eqref{RelativeEntropy} (cf. \cite{KPV:2021ea}). Accordingly, $\tilde{h}_{ij}$ is fixed by requiring that the first order geometric area variation needs to reproduce the area change induced by the horizon energy flux $\delta Q$, which is given by the relative entropy \eqref{RelativeEntropy}. Taking the ansatz
\begin{equation}\label{MetricPerturbation} 
\tilde{h}_{ij} = h_{ij} \left( 1-\varepsilon \, \alpha \int_{(-\infty,0)} U \langle :T_{ab}:\rangle_{\omega_\phi}\xi^a\xi^b dU\right), \end{equation}

\noindent for some proportionality constant $\alpha>0$, we use that $\sqrt{-\tilde{h}} = \left( 1-\varepsilon \, \alpha \int U \langle :T_{ab}:\rangle_{\omega_\phi}\xi^a\xi^b dU\right) \sqrt{-h}$ for the perturbation \eqref{MetricPerturbation} of the two-dimensional submanifold $\mathcal{S}$, in order to consistently verify that
\begin{align}
\delta A = \left. \frac{d A(\tilde{\mathcal{S}})}{d\varepsilon} \right\vert_{\varepsilon=0} &= - \alpha \int_{\mathcal{H}_B^\mathcal{R}} U \langle :T_{ab}:\rangle_{\omega_\phi}\xi^a\xi^b dU d\mathrm{vol}_\mathcal{S} \nonumber \\
&= - \alpha \int_{\mathcal{H}_B^\mathcal{R}} U (\partial_U \phi)^2 dU d\mathrm{vol}_\mathcal{S} \nonumber \\
&= \frac{\alpha}{2\pi} S^\mathrm{rel}\left(\omega_0 \Vert \omega_\phi \right).
\label{AreaVariationEntropicEnergy}
\end{align} 

On the other hand, the variation $\delta A$ of the horizon cross section surface area can be geometrically related to the focussing of null geodesics by using the expansion scalar $\theta$ of the (ingoing) null geodesic congruence on $\mathcal{H}_B^\mathcal{R}$ with tangent vector $\xi^a$, see \cite{Poisson:2004tool}. More precisely, it holds that \cite{Jacobson:1995eos,Poisson:2004tool}
\begin{equation}
\delta A = \int_{(-\infty,0)\times\mathcal{S}} \theta \, dU d\mathrm{vol}_\mathcal{S}.
\end{equation}

\noindent Moreover, considering the Raychaudhuri equation \cite{Poisson:2004tool}
\begin{equation}\label{RaychaudhuriEquation}
\frac{d\theta}{dU} = -\frac{\theta^2}{2} - \sigma_{ab} \, \sigma^{ab} + \omega_{ab}\,\omega^{ab} - R_{ab} \,\xi^a \xi^b,
\end{equation}

\noindent for null geodesic congruences, and using that on any bifurcate Killing horizon, the expansion scalar $\theta$, the shear tensor $\sigma^{ab}$, as well as the vorticity tensor $\omega^{ab}$ all vanish \cite{Wald:1984gr,Poisson:2004tool}, such that in a neighborhood of the approximate Killing horizon $\mathcal{H}_B$ we have $\theta^2 \approx 0$, $\sigma_{ab}\,\sigma^{ab} \approx 0$, and $\omega_{ab}\,\omega^{ab} \approx 0$, we find in analogy to Ref. \cite{Jacobson:1995eos} that
\begin{equation}
\theta \approx - U R_{ab}  \,\xi^a \xi^b,
\end{equation}

\noindent which leads us to
\begin{equation}\label{AreaVariationGeometricRicci}
\delta A = - \int_{(-\infty,0)\times\mathcal{S}} U R_{ab}  \,\xi^a \xi^b \, dU d\mathrm{vol}_\mathcal{S}.
\end{equation}

Identifying the area variations \eqref{AreaVariationEntropicEnergy} and \eqref{AreaVariationGeometricRicci} of the bifurcation surface $\mathcal{S}$, it follows immediately that 
\begin{equation}
\alpha \, \left\langle :T_{ab}: \right\rangle_{\omega_\phi} \, \xi^a \xi^b = R_{ab}  \,\xi^a \xi^b,
\end{equation}

\noindent which means that $\left\langle :T_{ab}: \right\rangle_{\omega_\phi}$ must be proportional to the Ricci tensor plus possibly some additional terms that vanish upon contraction with null vector fields, i.e.,
\begin{equation}
\alpha\, \left\langle :T_{ab}: \right\rangle_{\omega_\phi} = R_{ab} + N \,g_{ab},
\end{equation}

\noindent for a suitable coordinate function $N$ on $\mathcal{M}$. Given that due to local energy momentum conservation, see \cite{Wald:1977br,AL:1978ta,BD:1982qftcs}, it holds that $\nabla^a \left\langle :T_{ab}: \right\rangle_{\omega_\phi} = 0$, and hence \cite{Jacobson:1995eos}
\begin{equation}
N = -\frac{R}{2} + \Lambda,
\end{equation}

\noindent where $R$ denotes the Ricci curvature scalar and $\Lambda\in\mathbb{R}$ is an arbitrary constant, which shall be identified with the cosmological constant.

Altogether, this yields the semiclassical Einstein equations 
\begin{equation}\label{SemiclassicalEinsteinEquations}
R_{ab} - \frac{R}{2} g_{ab} + \Lambda g_{ab} = \alpha \,  \left\langle :T_{ab}: \right\rangle_{\omega_\phi},
\end{equation}

\noindent with arbitrary proportionality constant $\alpha$. In particular, if we assume in analogy to the Bekenstein-Hawking entropy-area relation, see \cite{Bekenstein:1973bhe,Hawking:1975rad,Hawking:1976bhe}, that the relative entropy is equal to one fourth of the area variation, then, the constant $\alpha$ naturally takes the value $8\pi$, coinciding with the standard proportionality factor of the Einstein equations. At last, we emphasize that the converse direction holds as well, namely that the semiclassical Einstein equations \eqref{SemiclassicalEinsteinEquations} imply the general relation $\delta A= 4 S^{\mathrm{rel}}$, reminiscent of the Bekenstein-Hawking entropy-area law.

\paragraph*{Conclusions.}
We illustrate a direct relation between the relative entropy between coherent states on approximate bifurcate Killing horizons, particularly local Rindler horizons, to the energy flux across the horizon, which is, in turn, assumed to be proportional to the variation of the horizon cross section surface area \cite{Jacobson:1995eos}. Analogously to \cite{Jacobson:1995eos}, this implies the semiclassical Einstein equations without further input. Thereby, we provide an extension of Jacobson's thermodynamic approach to a quantum field theoretical setting.

From the physical perspective, $S^{\mathrm{rel}}(\omega_0\Vert\omega_\phi)$ quantifies the information theoretic distinguishability between the vacuumlike reference state $\omega_0$ and its coherent excitation $\omega_\phi$ on the horizon algebra $\mathscr{W}_\mathcal{R}$. In the scaling limit theory on $\mathcal{H}_B^\mathcal{R}$, such coherent states represent the simplest model of infalling matter: The theory reduces to a conformal massless field, and the coherent excitation corresponds to a minimal deviation from the vacuum. 

Crucially, coherent state excitations reduce entanglement compared to the maximally entangled vacuum state. Our relative entropy calculation thus quantifies this entanglement deficit precisely, measuring how matter configurations differ in an informational sense from empty space. At the same time, the relative entropy depends directly on the energy content of the coherent state as seen with respect to the Killing flow on the horizon, thereby linking the difference in information induced by the coherent excitation to the excess of energy with respect to the vacuum state. Following the route of Ref. \cite{Jacobson:1995eos}, i.e., employing the proportionality relation $S^{\mathrm{rel}}=\tfrac{\delta A}{4}$, and requiring statistical consistency across all local Rindler horizons, necessarily leads us to the semiclassical Einstein equations. More generally, we expect this derivation to also extend to possibly noncoherent states; however, in the present letter, we restrict ourselves to coherent states due to the lack of explicit results for more general cases. Altogether, our findings indicate that the semiclassical Einstein equations, and thus local spacetime curvature, can be interpreted as arising from quantum information, namely the distinguishability between vacuum and excited states on local horizons.

To verify consistency, we may consider $\phi=0$, in which case no genuine excitation is present. Consequently, the relative entropy vanishes, so that $\omega_0$ and $\omega_\phi$ coincide, and the area variation vanishes, as well. Therefore, the corresponding null congruence remains unfocused, consistent with an exactly flat local spacetime geometry.

As a final remark, we emphasize that our analysis, much like Jacobson’s original argument \cite{Jacobson:1995eos}, should be understood as a leading order approximation whose validity relies on the accuracy of the local Minkowski approximation \eqref{LocalApproximation}. A fully rigorous treatment would require control of higher order corrections, e.g., by employing Riemannian normal coordinates. Although it is expected to be technically demanding, especially regarding the modular data, clarifying these corrections remains an important and promising direction for future research. \\

Looking further ahead, it is worth noting that recent advances express the quantum null energy condition in terms of the relative entropy using modular theory and half-sided modular inclusions, see \cite{HL:2025qnec,MTW:2022npa}. Given that our approach ties the relative entropy to the energy flux across local geometric horizons, it would be of interest to investigate how this relation aligns with those results.\\

\paragraph*{Acknowledgements.}
The authors thank Markus Fröb for very helpful discussions and suggestions, as well as Rainer Verch for particularly insightful and valuable comments. They also thank Bernard Kay for constructive feedback. A.M. is thankful for the stimulating atmosphere at the Causal Fermion Systems 2025 conference that lead to ideas implemented in this work.

\bibliography{bibliography}

\end{document}